\documentstyle[12pt,epsf]{article}
\newcommand {\beq}{\begin{equation}}
\newcommand {\eeq}{\end{equation}}
\newcommand {\beqa}{\begin{eqnarray}}
\newcommand {\eeqa}{\end{eqnarray}}
\newcommand {\beqan}{\begin{eqnarray*}}
\newcommand {\eeqan}{\end{eqnarray*}}
\newcommand {\n}{\nonumber \\}

\newcommand {\Romannumeral}[1]{\uppercase\expandafter{\romannumeral#1}}

\begin{document}
\setlength{\oddsidemargin}{0cm}
\setlength{\baselineskip}{7mm}  %7mm 

\begin{titlepage}
 \renewcommand{\thefootnote}{\fnsymbol{footnote}}
    \begin{normalsize}
     \begin{flushright}
                 KEK-TH-515\\
                 hep-ph/9703421\\
                 March 1997
~~\\
~~\\
~~\\
~~
     \end{flushright}
    \end{normalsize}
\vspace*{0cm}
    \begin{Large}
       \vspace{2cm}
       \begin{center}
         {\Large String-Scale Baryogenesis} \\
       \end{center}
    \end{Large}

  \vspace{1cm}

\begin{center}
           Hajime A{\sc oki}\footnote
           {
e-mail address : haoki@theory.kek.jp,
{}~JSPS research fellow.},
           Hikaru K{\sc awai}\footnote
           {
e-mail address : kawaih@theory.kek.jp}\\
      \vspace{1cm}
         {\it National Laboratory for High Energy Physics (KEK),}\\
               {\it Tsukuba, Ibaraki 305, Japan} \\
\end{center}

\vfill

\begin{abstract}
\noindent 
Baryogenesis scenarios at the string scale are considered.
The observed baryon to entropy ratio, $n_B /s \sim 10^{-10}$,
can be explained in these scenarios.
\end{abstract}
\vfill
\end{titlepage}
\vfil\eject

%\vspace{1cm}

\section{Introduction}
\setcounter{equation}{0}
\hspace*{\parindent}
Baryogenesis \cite{Kolb Turner} is one of the important problems 
in particle physics and cosmology.
Why is our universe made of matter, not anti-matter?
How do we explain the observed value of the ratio $n_B /s \sim 10^{-10}$,
where $n_{B}$ is the difference between the number density of baryons 
and that of anti-baryons, and $s$ is the entropy density?
Three ingredients are necessary to dynamically generate 
a nonzero $n_{B}$ 
from a baryon-symmetric initial state \cite{Sakharov}:\\
(1) baryon number nonconservation,\\
(2) violation of both C and CP invariance,\\
(3) departure from thermal equilibrium.\\

Baryogenesis scenarios at the electroweak scale have been studied recently
\cite{CKN}.
Baryon number conservation is violated at the electroweak scale {\em via} sphaleron 
processes.
However, it is difficult to generate the observed baryon to entropy 
ratio within the Minimal Standard Model (MSM).
First of all, CP violation coming only from the CKM phase is too small to 
explain the observed value, even if thermal plasma effects are taken into account
\cite{FS}.
Secondly, the electroweak phase transition should be a strong first-order 
phase transition in order to avoid the wash-out problem.
However, this requirement gives an upper bound on the Higgs boson mass
\cite{Shap}, 
which is already ruled out by LEP experiments.

Within the context of grand unified theories (GUT's) and the expanding universe
all three necessary conditions are satisfied.
However, GUT's also predict super-heavy magnetic monopoles which lead to a 
serious cosmological problem.
Another problem of GUT baryogenesis scenarios is that the baryon asymmetry
produced at the GUT scale is washed out by sphaleron processes. 
%\cite{Harvey Turner}.  

In this paper we will consider baryogenesis scenarios at the string scale
or the Planck scale and show how the observed baryon to entropy ratio can
be explained in these new scenarios.
Even if the non-SUSY non-GUT MSM describes the nature well above the 
electroweak scale,
it must be modified around the string scale or the Planck scale
due to gravitational effects.
Hence it is important to consider the baryogenesis scenarios at these scales. 

At the string and Planck scales, the three necessary conditions for 
baryogenesis are satisfied.
Let's consider string inspired models or 
effective theories with a cut-off at the string scale.
There is no reason to prohibit baryon- or lepton-number 
violating interactions in theories with a cut-off.
On the other hand, the MSM, 
which is required to be renormalizable and gauge invariant,
does not allow such interactions.
Sources of CP violation at the string scale can differ 
from those at the electroweak scale,
and other sources than the CKM phase are allowed at the string scale.

As for departure from thermal equilibrium, we consider two scenarios
which cause nonequilibrium distributions of matter.
The first uses the so called Hagedorn temperature \cite{Hagedorn,Hagedorn2}.
String theory has a limiting temperature, where the higher excited states of 
string theory are occupied.
The decay processes of these states will cause nonequilibrium distributions.
The other is the inflation scenario  \cite{inflation},
where inflaton decay processes cause nonequilibrium.

In this paper we will mainly consider the Hagedorn scenario.
Nonzero $n_{B}$ is generated during the decay of the higher excited states.
It is also generated after the decay 
since nonequilibrium distributions caused by the decay processes
are maintained 
until the rates for thermalization processes dominate the Hubble expansion rate. 

The resultant baryon to entropy ratio will not have suppression factors
since the theory has only one scale, the string scale or the Planck scale.
Hence, we expect the observed value is obtained in these scenarios.

In Section \ref{sec:model} we present a model and show
how it satisfies the three conditions for baryogenesis.
In Section \ref{sec:Boltzmann} we calculate the resultant lepton asymmetry
by considering Boltzmann equations 
and show that these scenarios can explain the observed baryon to entropy ratio.
The last section is devoted to conclusions and discussions.     
\vspace{1cm}

\section{A Model}
\label{sec:model}
\setcounter{equation}{0}
\hspace*{\parindent}
In this section we will present a model of string scale baryogenesis. 
There has been progress in the study of string models without 
SUSY or GUT recently \cite{Dienes}, which we think are interesting.
Hence, as an effective theory of string theory, 
we consider a model whose matter content is the same 
as that of the MSM.
For simplicity, we consider lower-dimensional operators.
%since higher dimensional ones give
%results suppression factors of powers of  $T_{H}/m_{st}$, where 
%$T_{H} \sim 5\times 10^{16}{\rm GeV}$ 
%and $m_{st} \sim 5 \times 10^{17}{\rm GeV}$ are 
%Hagedorn temperature and string scale, respectively.
Let's consider the  following model.
\beqa
{\cal L} &=& {\cal L}_{MSM} \n 
         & &  +\frac{1}{4} \frac{g_{st}^2}{m_{st}} h_{ij} \epsilon^{\alpha \beta}
               (\epsilon_{ab} \epsilon_{cd}+\epsilon_{ad} \epsilon_{cb})
            l_{\alpha}^{ia} \phi^b l_{\beta}^{jc} \phi^d
           +h.c.\n 
         & &  +\frac{1}{4} \frac{g_{st}^2}{m_{st}^2} \epsilon^{\alpha \beta}
            \epsilon^{\gamma \delta}
            [C_{ijkl}(\delta_{ac} \delta_{bd} + \delta_{ad} \delta_{bc})
             +C'_{ijkl}(\delta_{ac} \delta_{bd} - \delta_{ad} \delta_{bc}) 
             +C''_{ijkl} \epsilon_{ab} \epsilon_{cd}] \n 
 & & \times l_{\alpha}^{ia} l_{\beta}^{jb} l_{\gamma}^{kc*} l_{\delta}^{ld*},
\label{eq:model}
\eeqa
where $l$'s are the lepton doublets and $\phi$ is the Higgs doublet.
The string coupling constant and the string scale are $g_{st}$ and $m_{st}$,
respectively. 
Also, $i,j,k$ and $l$ are generational indices,
$\alpha,\beta,\gamma$ and $\delta$ are spinorial indices, 
and $a,b,c$ and $d$ are $SU(2)_L$ gauge-group indices.

The first term violates lepton number conservation since it is a Majorana-type
coupling.
The second term is included to incorporate CP violation.
Through a unitary transformation
the coefficient of the first term, $h_{ij}$, can be rotated to real
diagonal form.
However, if the second term is present we cannot guarantee
that both terms can be rotated to real form simultaneously.  
With two or more generations, this model violates CP invariance.
Henceforth, we will consider exactly two generations for simplicity .

In the remainder of this section we will speculate on the 
departure from thermal equilibrium.
First of all, let's consider how the universe would have been 
in the context of string theory if thermal equilibrium had been maintained
\cite{Hagedorn, Hagedorn2}.
Since the density of states is exponentially rising in string theory,
it has a limiting temperature, the Hagedorn temperature,
\beq
T_{H}=\left\{ \begin{array}{ll} 
                 m_{st}/2 \sqrt{2}\pi \sim 5.93\times 10^{16} {\rm GeV} 
               & ({\rm type \hspace{2mm} II}),\\
                 m_{st}/(2+\sqrt{2})\pi \sim 4.92\times 10^{16} {\rm GeV}
               & ({\rm heterotic \hspace{2mm} string}).
              \end{array}       
       \right.
\eeq
While the universe is contracting, the energy density increases 
but the temperature remains just below the Hagedorn temperature.
When the energy density is low compared to the string scale,
%the single string density of states is  
%\beq
%\omega(\epsilon) = V \frac{\exp(\epsilon / T_{H})}{\epsilon^{D/2+1}},
%\eeq            
%where V is the volume of the system.
matter is  composed of particles, or the excitations of short strings  
whose lengths are of the order of $m_{st}^{-1}$.
However, the high-energy limit of 
the single-string density of states is found to be \cite{Hagedorn2}
\beq
\omega(\epsilon) = \frac{\exp(\epsilon / T_{H})}{\epsilon}.
\eeq
When we consider the microcanonical ensemble
it turns out that long strings traverse the entire volume of the universe
in sufficiently high energy density.

Next we consider whether the thermal equilibrium is actually realized
by comparing the rates for thermalization processes with  the Hubble expansion
rate.
When matter is composed of particles, 
the rates for thermalization processes are
$\Gamma_{{\rm th}} \sim g_{*}\alpha_{st}^{2}T$, and
the Hubble expansion rate is $H = 1.66 g_{*}^{1/2}T^2/m_{pl}$, where
$m_{pl}$ and $g_{*}$ are the Planck mass and the number of matter species,
respectively, 
and $\alpha_{st} = g_{st}^2/4\pi$. 
Hence, above the temperature
$T_{{\rm eq}} \sim g_{*}^{1/2}\alpha_{st}^{2}m_{pl}/1.66 
\sim 3.8 \times 10^{16} {\rm GeV}$,
or above the energy density 
$\rho_{{\rm eq}} \sim (1.6 \times 10^{17}{\rm GeV})^4$,
the thermalization processes are too slow to maintain equilibrium distributions.
However, in the long-string phase, the rates for thermalization processes
are proportional to 
the densities of the string-bits, $E/V$,
%the probabilities of intersections among strings,
%$l^2/V \sim (E/\ln E)^2/V$, where $l$ is the characteristic length of the 
%strings, and 
where $E$ and $V$ are the total energy and the volume of the universe, 
respectively.
Since the Hubble expansion rate is proportional to 
the square root of the energy density, 
$\sqrt{E/V}$,
the rates for thermalization processes will dominate
the Hubble rate for  sufficiently high energy density.
Therefore, there will be a critical energy density, $\rho_{*}$,
above which equilibrium distributions are realized.
Below $\rho_{*}$ interactions freeze out and matter distributions 
are merely affected by the expansion of the universe 
and depart from thermal equilibrium distributions.

Decay processes of higher excited states begin when the energy density decreases
to $\rho_{{\rm decay}} \sim (3.7 \times 10^{17}{\rm GeV})^4$,
where the decay rates $\sim \alpha_{st} m_{st}$ dominate the Hubble expansion rate.
These processes are not adiabatic since the matter distributions have departed
from equilibrium ones.

During the decay processes of higher excited states, 
the baryon asymmetry as well as entropy is generated.
We can make a rough estimate,
\beq
n_{B}/s \sim \alpha_{st}^2 \sim 10^{-3}, \label{eq:roughdecay}
\eeq
since the first nontrivial contribution to CP violation comes from
the interference of the lowest-order diagrams and the one-loop corrections.

However, if many processes occur at $\rho =\rho_{{\rm decay}}$,
some cancellations among the decay processes can decrease 
the above result.
These cancellations might be possible 
if many excited states are taken into account
since ten-dimensional superstring theory has no CP violation originally.
They might be explained also since CPT invariance and unitarity
assure the following relation: 
$ \sum_{X} \Gamma (X \rightarrow b) = \sum_{X} \Gamma (X \rightarrow \bar{b})$.

We cannot make a precise estimate since we don't know the dynamics of the decay
processes in detail.
Hence let's consider the following case:
Nonzero $n_{B}$ is generated after the decay while 
it is not generated during the decay due to exact cancellation. 
By considering this case we can give a lower bound on $n_{B}/s$. 
We assume that the following matter distributions are caused by the 
decay processes:
\beq
n_{\phi} = n_{\phi^*} \neq n_{l} = n_{l^*} \hspace{1cm} (T=T_{H}),
\label{eq:initial}
\eeq    
where $n_{\phi}$ and $n_{l}$ are the number densities of Higgs bosons and lepton
doublets, respectively.
These nonequilibrium distributions are maintained 
until the temperature of the universe decreases to 
$T_{{\rm eq}}$.
During this epoch a nonzero $n_{L}$ is generated.
This nonzero $n_{L}$ will be converted into a nonzero $n_{B}$ 
of the same order of magnitude 
{\em via} sphaleron processes.
In the next section, we will estimate the resultant lepton asymmetry
by using the nonequilibrium distribution of Eqn.(\ref{eq:initial}) as an 
initial condition.
 
Finally, we will make a brief comment on another scenario which causes 
nonequilibrium distributions, {\em i.e.}, inflation.
While the inflaton decays, a nonzero $n_{B}$ as well as entropy can be generated
\cite{inflation}.
Even if a nonzero $n_{B}$ is not generated during the decay processes,
some nonequilibrium distributions like those of Eqn.(\ref{eq:initial}) 
are generated.
Then a nonzero $n_{B}$ is generated after the inflaton decays.

\vspace{1cm}

\section{Boltzmann Equations} \label{sec:Boltzmann}
\begin{figure}
\begin{center}
\leavevmode
\epsfbox{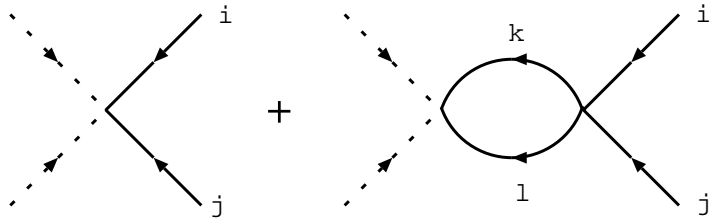}
\caption{Diagrams which contribute to leptogenesis.
The first nontrivial contributions come from the interference between
tree-level amplitudes and one-loop corrections.
The indices, $i,j,k$ and $l$, represent the generations.}
\label{fig:lepton}
\end{center}
\vspace{1cm}
\end{figure}

\begin{figure}
\begin{center}
\leavevmode
\epsfbox{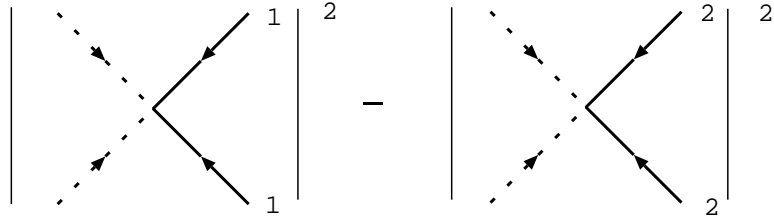}
\caption{Processes which produce 
         a different  number density for generation 1 and generation 2. }
\label{fig:generations}
\end{center}
\vspace{1cm}
\end{figure}

\setcounter{equation}{0}
\hspace*{\parindent}
In this section we will calculate the resultant lepton asymmetry  by
using the model of Eqn.(\ref{eq:model}) and the nonequilibrium distributions 
of Eqn.(\ref{eq:initial}). 
We consider the processes, $ll \leftrightarrow \phi^{*}\phi^{*}$ and 
processes related by CP conjugation.
The first nonzero contributions to the generation of nonzero $n_{L}$
come from the 
interference terms of the tree-level amplitudes and the one-loop corrections 
shown in Fig.\ref{fig:lepton}.
They are proportional to $\Im (\sum_{k,l} h_{ij}^* C_{ijkl} h_{kl}) \Im (I)$,
where $I$ is the factor coming from the loop integrations.
However, they vanish if we naively sum over the indices for generations $i,j$,
since $C_{ijkl}^* = C_{klij}$ as is evident form the Lagrangian of 
Eqn.(\ref{eq:model}).
Hence, we consider the processes shown in Fig.\ref{fig:generations}
in order to produce a different number density
for generation 1 and generation 2.

For simplicity, we will make the following assumption:
the distributions for matter remain  near equilibrium, 
$\rho \sim \exp(\frac{E-\mu}{T})$ and $\mu \ll T$.
Thus the Boltzmann equations for the above processes are as follows:
\beqa
&&\dot{Y_l}+\Gamma_{{\rm th}} (Y_l -1) = 0, \n 
&&\dot{Y_{\phi}}+\Gamma_{{\rm th}} (Y_{\phi} -1) = 0, \label{eq:Boltzmann0}\\ 
&&\dot{Y_{1-2}}+\Gamma_{{\rm th}} Y_{1-2} = 
  \frac{3}{\pi}(\frac{\alpha_{st}^2}{m_{st}})^2 (h_{11}^2-h_{22}^2)
T^3(Y_{\phi}^2-Y_{l}^2),\label{eq:Boltzmann1}\\ 
&& \dot{Y_{L}} 
= -\frac{72}{\pi} \frac{\alpha_{st}^3}{m_{st}^4}h_{11}h_{22}\Im(C_{1122})
T^5(Y_{l_1}^2-Y_{l_2}^2), \label{eq:Boltzmann2}
\eeqa
where $Y_i = n_i/n^{({\rm eq})}$.
$Y_{1-2}=Y_{l_1}-Y_{l_2}$, $Y_{L}=Y_{l_1}+Y_{l_2}-(Y_{l_1^*}+Y_{l_2^*})$.
$\Gamma_{{\rm th}} \approx g_* \alpha_{st}^2 T$ is the rate for thermalization.
We took the convention where the coefficient of the Majorana-type interactions,
$h_{ij}$, is real diagonal form.

Eqns.(\ref{eq:Boltzmann0}) represent the thermalization
processes which reduce the nonequilibrium distributions of Eqn.(\ref{eq:initial})
imposed as an initial condition to the equilibrium ones. 
Eqn.(\ref{eq:Boltzmann1}) represents the processes which produce
the difference in the number densities between the generations.
Finally, the third equation, Eqn.(\ref{eq:Boltzmann2}), represents 
the production of a nonzero $n_{L}$.
The right-hand sides of Eqns.(\ref{eq:Boltzmann1}) and (\ref{eq:Boltzmann2})
are given by calculating the amplitudes shown in Fig.\ref{fig:generations}
and Fig.\ref{fig:lepton}, respectively, 
and performing the phase space integrations.

The Boltzmann equations are integrated to give
\beqa
Y_{L}&=& -\frac{432}{\pi^2} \frac{\alpha_{st}^{5}}{m_{st}^6}
        (\frac{m_{pl}}{1.66g_*^{1/2}})^2
        (h_{11}^2-h_{22}^2)h_{11}h_{22} \Im (C_{1122}) \n
     & & \times \int_{T_{H}}^T dT' T'^2 Y_l(T') 
         \exp(-T_{{\rm eq}}/T') \n
     & & \times \int_{T_{H}}^{T'} dT'' 
         \exp(T_{{\rm eq}}/T'')
         (Y_{\phi}(T'')^2-Y_{l}(T'')^2),
\eeqa
where 
\beq
Y_i(T) = 1+(Y_i(T_{H})-1) 
         \exp (-T_{{\rm eq}}/T+T_{{\rm eq}}/T_{H}),
\eeq
for $i=l,\phi$, and 
$T_{{\rm eq}}=g_*^{1/2}\alpha_{st}^2 m_{pl}/1.66 
\sim 3.75 \times 10^{16} {\rm GeV}$ is the temperature below which 
thermalization processes begin.
Note that the Hubble expansion rate appears when we change the variable
from time to temperature.
The final result is as follows:
\beqa
Y_{L}&=& -\frac{36}{\pi^2} \alpha_{st}^{5}
         (\frac{m_{pl}}{1.66g_*^{1/2} m_{st}})^2 (\frac{T_{H}}{m_{st}})^4 J \\
     &\approx & 2.7 \times 10^{-12} J,\label{eq:semifinal}
\eeqa
where
\beqa
J &=& (h_{11}^2-h_{22}^2)h_{11}h_{22}\Im(C_{1122})K, \label{eq:J}\\
K &=& 12 (T_{{\rm eq}}/T_{H})^4 
         \int_{T_{{\rm eq}}/T_{H}}^{\infty}dz z^{-4}(\exp(-z)+A_{l}\exp(-2z))\n
  & &       \times \int_{T_{{\rm eq}}/T_{H}}^z dz' z'^{-2}
          (2(A_{\phi}-A_{l})+(A_{\phi}^2-A_{l}^2)\exp(-z'))\label{eq:integral}\\
  &\approx&  0.538(A_{\phi}-A_{l})+0.173A_{l}(A_{\phi}-A_{l})\n
  & &          +0.108(A_{\phi}^2-A_{l}^2)+0.0355A_{l}(A_{\phi}^2-A_{l}^2) 
             \label{eq:integral2},\\ 
A_i &=& (Y_i (T_{H})-1)  \exp (T_{{\rm eq}}/T_{H}). \\
\eeqa
In the estimations of Eqns.(\ref{eq:semifinal}) and (\ref{eq:integral2})
we used the following values:
$\alpha_{st} = 1/45$, $g_{*}=106.75$, $m_{pl}=1.22\times10^{19}{\rm GeV}$,
$m_{st}=5.27\times10^{17}{\rm GeV}$ and $T_{H}=4.92\times10^{16}{\rm GeV}$.
When $T_{{\rm eq}}/T_{H} \ll 1$, thermalization processes can be neglected, and
the integration in Eqn.(\ref{eq:integral}) approaches  
\beq
K \rightarrow (1+A_{l})[2(A_{\phi}-A_{l})+A_{\phi}^2-A_{l}^2]
\hspace{1cm}(T_{{\rm eq}}/T_{H} \ll 1).
\eeq
The result in Eqn.(\ref{eq:integral2}) was calculated numerically
with the value  $T_{{\rm eq}}/T_{H} \approx 0.763$.
Note that the result has no suppression as far as  $T_{{\rm eq}}/T_{H}$
is of the order one. 

The lepton to entropy ratio is 
\beq
n_{L}/s \approx Y_{L}/g_{*} \approx 2.6\times10^{-14} J,
\label{eq:final}
\eeq
where $J$ is given by Eqn.(\ref{eq:J}) and its value can be a few thousand
if $h_{ij}$ and $C_{ijkl}$ are about five.
This nonzero $n_{L}$ will be converted into nonzero $n_{B}$ of the  
same order of magnitude {\em via} sphaleron processes.
Therefore, the observed baryon asymmetry can be generated after the 
decay processes of higher excited states of string theory. 

\vspace{1cm}

\section{Conclusions and Discussions}
\setcounter{equation}{0}
\hspace*{\parindent}
In this paper we consider baryogenesis scenarios at the string scale 
or the Planck scale.
At these scales the three necessary conditions for baryogenesis are satisfied.
We have shown that the observed baryon asymmetry 
can be generated after the decay processes
of higher excited states of string theory.
%by considering the  string inspired model 
%whose matter contents are same as those in MSM.
If we take into account the baryogenesis during the decay processes,
we will obtain a larger value for the baryon to entropy ratio.
Too large a value could be diluted afterwards
by considering entropy generation in the confinement-deconfinement
phase transition, for example.
Therefore, the observed baryon to entropy ratio can be explained 
in these scenarios.

%We should also identify the origins of lepton number violation and CP violation
%in the model (\ref{eq:model}) in some string models, for example,
%four-dimensional heterotic string model.

\begin{figure}
\begin{center}
\leavevmode
\epsfbox{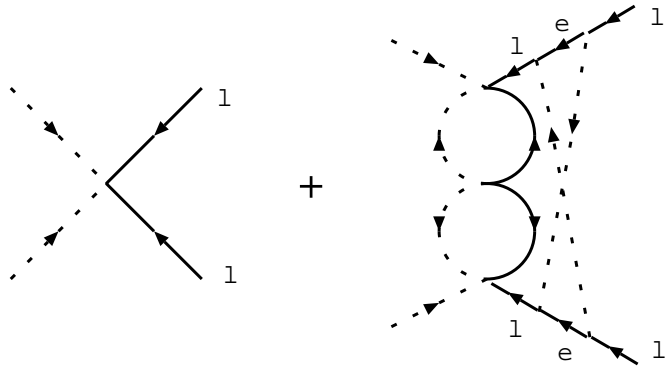}
\caption{Diagrams which contribute to leptogenesis 
by means of Yukawa couplings and Majorana-type couplings.
The left-handed lepton doublets and the right-handed lepton singlets are
$l$ and $e$, respectively.}
\label{fig:yukawa}
\end{center}
\vspace{1cm}
\end{figure}

Finally we give some comments on the model of Eqn.(\ref{eq:model}).
We have considered Majorana-type interactions plus
four-fermion interactions other than the MSM
in order to introduce CP violation.
It seems that only Majorana-type interactions would be enough since 
there are Yukawa couplings in the MSM already.
The Yukawa plus Majorana-type interactions,
\beq
y_{ij} e^{*\alpha i}l_{\alpha}^{aj} \phi^*_a
+\frac{1}{4} \frac{g_{st}^2}{m_{st}} h_{ij}
 \epsilon^{\alpha \beta}
(\epsilon_{ab} \epsilon_{cd}+\epsilon_{ad} \epsilon_{cb})
 l_{\alpha}^{ia} \phi^b l_{\beta}^{jc} \phi^d
+h.c.,
\eeq
would also work,
since, after the Yukawa coupling is brought to the form of a  real diagonal 
matrix {\em via} a unitary 
transformation, no degrees of freedom remain to insure a
real Majorana-type coupling.
However, because the Yukawa coupling is small, the resultant lepton 
asymmetry is too small to explain the observed value of $n_{B}/s$.
Indeed, 
a nonzero $n_{L}$ is generated, for example, 
through the processes shown in Fig.\ref{fig:yukawa}.
The result is of the order,
\beqa
n_L/s &\sim& (h\frac{\alpha_{st}}{m_{st}})^{4} y^4 
                      \frac{m_{pl}}{1.66 g_{*}^{1/2}}T_{H}^3 \frac{1}{g_{*}} \n
               &\sim& 6.7 h^4 \times 10^{-21}. 
\eeqa
Here $h$ and $y$ are characteristic values for $h_{ij}$ and $y_{ij}$, 
respectively.
We think $h$ is of the order one, and we used 
the Yukawa coupling of the tau lepton for $y$.

Finally, the Majorana-type terms in Eqn.(\ref{eq:model}) give neutrino masses
of the order of
\beq
m_{\nu} \sim h \frac{g_{st}^2}{m_{st}} v^2 
         \sim 3.2 h \times 10^{-5}\hspace{2mm}{\rm eV}.
\eeq
This value is consistent with solar neutrino experiments 
if we consider the vacuum oscillation or 
take into consideration the magnetic field in the sun \cite{Takasugi}.

\vspace{1cm}

\begin{center} \begin{large}
Acknowledgements
\end{large} \end{center}

We would like to thank K. Funakubo, K. Inoue and Y. Okada for useful discussions.
We are also grateful to R. Szalapski for carefully reading the manuscript.


\newpage

\begin{thebibliography}{99}
\bibitem{Kolb Turner}M. Yoshimura, Phys. Rev. Lett. {\bf 41} (1978) 281;\\
                     S. Weinberg, Phys. Rev. Lett. {\bf 42} (1979) 850;\\ 
                     For reviews, see: E. W. Kolb and M. S. Turner, 
                     Ann. Rev. Nucl. Part. Sci. {\bf 33} (1983) 645;\\
                     E .W. Kolb and M. S. Turner, {\it The Early Universe},
                     (Addison Wesley, California, 1990).
\bibitem{Sakharov}A. D. Sakharov, JETP Lett. {\bf 5} (1967) 24.
\bibitem{CKN}For reviews, see: A. G. Cohen, D. B. Kaplan and A. E. Nelson,
             Ann. Rev. Nucl. Part. Sci. {\bf 43} (1993) 27;\\
             K. Funakubo, Prog. Theor. Phys. {\bf 96} (1996) 475.
\bibitem{FS} G. R. Farrar and M. E. Shaposhnikov, Phys. Rev. {\bf D50}
             (1994) 774;\\
             M. B. Gavela, P. Hernandez, J. Orloff, O. P\`{e}ne and C. Quimbay,
             Nucl. Phys. {\bf B430} (1994) 382;\\
             P. Huet and E. Sather, Phys. Rev. {\bf D51} (1995) 379.
\bibitem{Shap} M. E. Shaposhnikov, Nucl. Phys. {\bf B287} (1987) 757.
%\bibitem{Harvey Turner}M. Fukugita and T. Yanagida, Phys. Rev. {\bf D42}
%                       (1990) 1285.\\
%                       J. A. Harvey and M. S. Turner, Phys. Rev. {\bf D42} 
%                       (1990) 3344.
\bibitem{Hagedorn}R. Hagedorn, Nuovo Cim. Suppl. {\bf 3} (1965) 147;\\
                  K. Huang and S. Weinberg, Phys. Rev. Lett. {\bf 25} (1970)
                  895;\\
                  J. J. Atick and E. Witten, Nucl. Phys. {\bf B310} (1988) 219.
\bibitem{Hagedorn2}R. Brandenberger and C. Vafa, Nucl. Phys. {\bf B316} 
                   (1989) 391; \\
                   D. A. Lowe and L. Thorlacius, Phys. Rev. {\bf D51} (1995) 665.
\bibitem{inflation}A. D. Dolgov and A. D. Linde, Phys. Lett. 
                   {\bf 116B} (1982) 329;\\
%                   A. Albrecht, P. J. Steinhardt, M. S. Turner and F. Wilczek,
%                   Phys. Rev. Lett. {\bf 48} (1982) 1437;\\
                   L. F. Abbott, E. Farhi and M. B. Wise, Phys. Lett. 
                   {\bf 117B} (1982) 29.
\bibitem{Dienes}K. R. Dienes, {\it String Theory and the Path to Unification},
                hep-th/9602045. 
\bibitem{Takasugi}T. Kubota, T. Kurimoto and E. Takasugi, Phys. Rev. {\bf D49}
                  (1994) 2462.
\end{thebibliography}
\end{document}